\documentclass[acmsmall]{acmart}
\usepackage{multirow}
\pdfoutput=1

\AtBeginDocument{%
	\providecommand\BibTeX{{%
			\normalfont B\kern-0.5em{\scshape i\kern-0.25em b}\kern-0.8em\TeX}}}

\renewcommand\footnotetextcopyrightpermission[1]{} % removes footnote with conference information in first column
\newcommand{\bem}[1]{\textbf{\emph{#1}}}

\begin{document}
	\title{A Note on Cryptographic Algorithms for Private Data Analysis in Contact Tracing Applications}
	
	\author{Rajan M A}
	\email{rajan.ma@tcs.com}
	\affiliation{%
		\institution{TCS Research}
		\city{Bangalore}
		\country{India}}
	
	\author{Manish Shukla}
	\email{mani.shukla@tcs.com}
	\affiliation{%
		\institution{TCS Research}
		\city{Pune}
		\country{India}}
	
	\author{Sachin Lodha}
	\email{sachin.lodha@tcs.com}
	\affiliation{%
		\institution{TCS Research}
		\city{Pune}
		\country{India}}
	
\begin{abstract}
	\bem{Abstract.} \emph{Contact tracing is an important measure to counter the COVID-19 pandemic. In the early phase, many countries employed manual contact tracing to contain the rate of disease spread, however it has many issues. The manual approach is cumbersome, time consuming and also requires active participation of a large number of people to realize it. In order to overcome these drawbacks, digital contact tracing has been proposed that typically involves deploying a contact tracing application on people's mobile devices which can track their movements and close social interactions. While studies suggest that digital contact tracing is more effective than manual contact tracing, it has been observed that higher adoption rates of the contact tracing app may result in a better controlled epidemic. This also increases the confidence in the accuracy of the collected data and the subsequent analytics.} 
	
	\emph{One key reason for low adoption rate of contact tracing applications is the concern about individual privacy. In fact, several studies report that contact tracing applications deployed in multiple countries are not privacy friendly and have potential to be used for mass surveillance by the concerned governments. Hence, privacy respecting contact tracing application is the need of the hour that can lead to highly effective, efficient contact tracing.}
	
	\emph{As part of this study, we focus on various cryptographic techniques that can help in addressing the Private Set Intersection problem which lies at the heart of privacy respecting contact tracing. We analyze the computation and communication complexities of these techniques under the typical client-server architecture utilized by contact tracing applications. Further we evaluate those computation and communication complexity expressions for India scenario and thus identify cryptographic techniques that can be more suitably deployed there.}
	
\end{abstract}

\maketitle

\pagestyle{empty}

\section{Introduction}
	In the current COVID-19 crisis, contact tracing mobile applications is the need of the hour to contain the pandemic, and hence privacy preservation becomes a core area of significance and relevance. As more people start using these applications, it will become easier for individuals, the government and the society at large to analyse the spread and achieve traceability of the affected people, and thereby render support. Through the app, individuals can be alerted if they had visited the places that were visited by a COVID-19 affected person around the same time. Similarly, it will help the government to identify the affected people and quarantine them, so that the rate of disease spread can be contained. Therefore, an extensive and effective use of the App can help to contain the disease and thereby help to reduce the causalities. It can also enable a reduction of the lockdown duration, thereby expediting the resumption of economic activities post lockdown and thus helping the society. One of the main reasons why many people are reluctant to use contact tracing applications include their privacy concerns, especially pertaining to the individual's location and health data, as the former can reveal a person's behaviour and disclose relationships, and the latter can have adverse impact on his/her social life. To address this, apps such as MIT's SafePath from USA \cite{raskar2020apps}, TraceTogether from Singapore, Hamagen (The Shield) from Israel, Aarogya Setu from India, etc. have been proposed and deployed with a focus on user centric privacy along with contact tracing \cite{shukla2020privacy}. Further, it is observed that higher adoption rates of a contact tracing app may result in a better controlled epidemic or pandemic outbreak \cite{yasaka2020peer}. In Indian context, it is proposed that at least 60-70\% of the population is required to use the contact tracing application actively, in order to have confidence in the data accuracy of contact tracing and the related risk calculation\footnote{https://theprint.in/opinion/india-can-prevent-long-term-lockdown-contain-covid-if-90-people-download-aarogya-setu/401116/}.
	
	Further OpenMined blog post \cite{openminded2020privacy} has a discussion on the maximisation of privacy without degrading the effectiveness of COVID-19 apps. They have done a detailed analysis of possible use cases from the perspective of both the citizens as well as the government health services. The blog post further describes techniques that can be used to design privacy enabled apps for citizen centric use cases. One of the main enablers for these kinds of apps is the Private Set Intersection (PSI) technique \cite{berke2020assessing}. In this work, we describe some such techniques that are based on well-studied cryptographic approaches that in turn facilitate flexible client-server mode of operations for such mobile applications.
	
	\section{Setup}
	As part of the setup, we assume that there is a server that has the GPS trails of COVID-19 affected persons and it can provide redacted data (by filtering out the sensitive and personally identifiable information such as residence/office location etc.) of these data sets to the client application (mobile app of the client side contact tracing app) which is running on the user's mobile phone in an encrypted manner. Here a person who is using the contact tracing client app in his/her mobile phone that has a list of GPS trails of the places he/she has visited. He/She can query the server through his/her client app to know whether he/she has crossed the places where a COVID-19 affected person visited at the same time or within a specified time window (as COVID-19 virus can survive for several hours) in a privacy preserving manner. Accordingly, a risk score can be computed. This problem can be abstracted as Private Set Intersection (PSI) problem. Informally, in PSI, each party has got a set of elements and they want to compute the intersection of these two sets in a privacy preserving manner. Here, both the parties can learn only the intersection and it is computationally very hard for them to know any details about the other elements of each other's sets. Generally, PSI protocols can be designed under three settings: (i) Pull model, (ii) Push model, and (iii) Hybrid model. We explain these models in the remainder of the section. 
	
	For analysis purpose, we assume that for the country of size M, the size of the server data will be $O(M)$ and if $O(M)$ people download the data from the server, then the data consumption will be $O(C(M^2))$. Here $C$ represents communication complexity of the given PSI protocol. It depends on cardinalities of the two sets, encryption scheme used in the protocol, etc. For the sake of simplicity, we assume that it depends on cardinalities of the two sets $M$ and $N$. Let $O(M)$ and $O(N)$ be the size of the set of GPS trails of the server and the client respectively.
	
	\subsection{Pull Model}
	The client pulls the GPS trails of the COVID-19 affected persons from the server and computes the intersection of his/her GPS trail set and the server's set in a privacy preserving manner. Note that, all the computations need to be performed at the client side and client app also needs to download the entire data set from the server. Thus, communication and computational complexities for the client are $O(C(M))$ and $O(F(M,N))$ respectively. Here $F$ is the computational complexity of a given PSI protocol, which depends on cardinalities of the two sets ($M$ and $N$), key length, etc. For the sake of simplicity, we assume that $F$ is a function of $M$ and $N$.
	
	\begin{itemize}
		\item \emph{Advantage for the Client}. Data privacy of the client is ensured, as the server cannot determine anything about the client's data.
		\item \emph{Disadvantage for the Client}. All the computations related to the PSI need to be performed locally and hence it consumes battery power.
		\item \emph{Advantage for the Server}. Less computation overhead, as server does not need to compute PSI.
		\item \emph{Disadvantage for the Server}. Client can carry out some inference attacks by brute forcing.			
	\end{itemize}
	
	\subsection{Push Model}
	The client can send his/her GPS trails to the server in an encrypted manner. The server computes the PSI and shares the result with the client. Thus, communication and computational complexity for the client and server are $O(C(N))/O(F(N))$ and $O(C(N))/O(F(MN))$ respectively.
	\begin{itemize}
		\item \emph{Advantage for the Client}. Client application will be lightweight and therefore, it can encourage more people to use the application.
		\item \emph{Advantage for the Server}. There is no need for the server to send its data to the client and hence it can prevent inference attacks from the client.
		\item \emph{Disadvantage for the Server}. Server has more computation overhead, as it needs to compute PSI.			
	\end{itemize}

	\subsection{Hybrid Model}
	In the hybrid approach, both the client and the server need to do computations for many rounds (at least two) along with data exchanges. Thus, the client and the server need to do both push and pull of the data from each other along with some computations. In most of the techniques such as RSA, etc., which are used to solve PSI, communication and computation complexity for the server is more than that of the client. For example, in case of RSA based PSI implementation, communication and computation complexity for the server and the client are $(O(C(M+N)),O(F(M+N)))$ and $(O(C(N)),O(F(N)+MN))$ respectively. Note that since, $M >> N$, the computation overhead of the server $(O(F(M+N)))$ is greater than that of the client $(O(F(N)+MN))$. To start with, for the sake of completeness, we discuss the simplest PSI scheme based on hash (na\"ive) approach followed by different cryptographic approaches.
	
	\subsubsection{Hash Based (Na\"ive) Approach}
	Here either the server or the client can send their data set to each other and compute the intersection and they can share the result with each other. From the user's data privacy perspective for COVID-19 scenario, the pull model is preferred (at the cost of higher computation and communication complexities), while the push model is not suitable as it leaks the client's data to the server. For the pull model, the client needs to perform $O(MN)$ comparisons to determine intersection. In case of the pull model, the client can learn the server's data through brute force.
	
	\section{Cryptographic Approaches to Compute PSI} \label{sec:schemes}
	In this section, we discuss different classes of cryptographic techniques to solve PSI. Throughout this analysis, we assume that set of GPS trails consists of an ordered tuple which includes latitude, longitude and timestamp. For the analysis purpose, let latitude, longitude and timestamp be concatenated and the concatenated data be hashed. The output of the hash of these concatenated data can be considered as elements of the set of GPS trails. On these sets, PSI can be performed. Let $X=\lbrace x_1,x_2,\dots,x_m \rbrace$ be the set of GPS trails of COVID-19 affected persons maintained by the server and $Y=\lbrace y_1,y_2,\dots,y_n \rbrace$ be the set of GPS trails of the user of the contact tracing app stored by the client app running on the user's mobile phone. Here, each $x_i$ and $y_i$ are obtained by hashing the concatenation of latitude, longitude and timestamp related to COVID-19 affected person and app user respectively. We assume that $m \gg n$. Further, most of these PSI schemes assume semi-honest adversaries (wherein both the parties are curious, but do not deviate from the protocol execution). Therefore, there is computationally less overhead when compared to those protocols that handle malicious adversaries. Since these apps need to perform PSI on large data set, very often protocols that support semi-honest adversaries are suitable.
	
	There are broadly three approaches: (i) Public key crypto based (ii) Circuit based (iii) Oblivious Transfer (OT) based, under which PSI schemes are designed. Circuit based and OT based schemes are generally not suitable for large sized sets ($> 2^{30}$), where the client system is deployed on a handheld device/mobile phone, as these schemes are computationally complex and the circuit generation needs to be done every time for any new queries \cite{kiss2017private}. For example, for Yao-SCS (Sort Compare Shuffle) based PSI scheme \cite{huang2012private,pinkas2018scalable}, computation and communication complexities are $O((12m\sigma\log m + 3m\sigma))$ symmetric operations and $O(6m\kappa\sigma\log m + 2m\kappa\sigma)$ bits respectively. Here $\sigma$ and $\kappa$ are length of the input (at least 256 bits) and security parameter of the symmetric encryption respectively. The higher the cardinality of the server data set, the higher will be the computation and communication complexities.
	
	Since contact tracing applications are installed on mobile phones, we now discuss only those PSI schemes that are based on public key crypto systems, which are suitable to compute $X \cap Y$ and analyse their suitability from the perspective of the above models.
	
	%\begin{itemize} \item \textbf{\emph{Diffie-Hellman Approach [4]}}. 
	\subsection{Diffie-Hellman Approach}
\begin{figure}[h]
			\centering \includegraphics[trim=0cm 12.5cm 6cm 0cm, clip=true,scale=0.45]{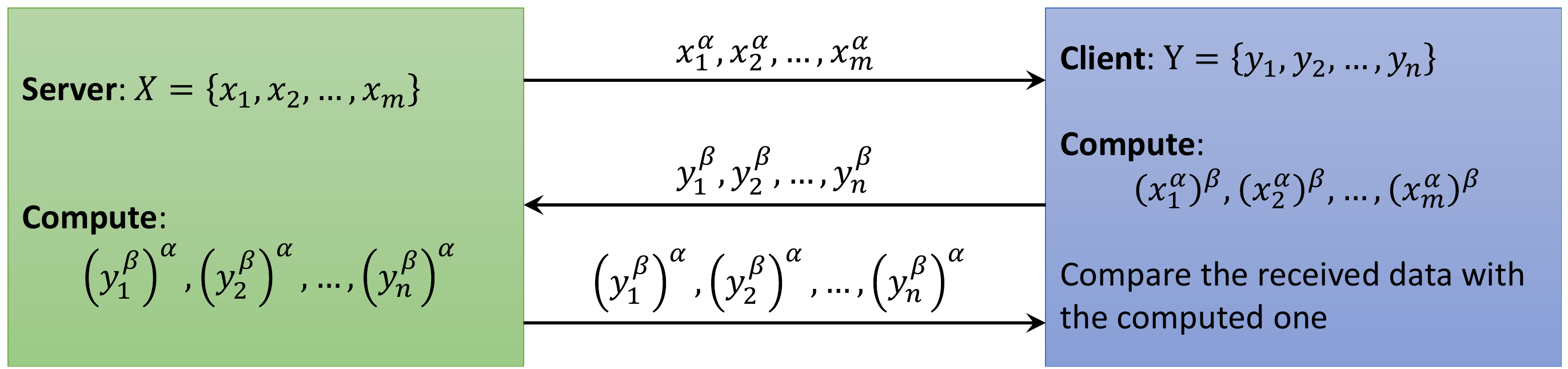} %trim=l b r t
			\caption{PSI Scheme Based on Diffe-Hellman Approach} \label{fig:diffe}
			\Description{}
		\end{figure}

		Security of this algorithm is based on DDH (Decisional Diffie-Hellman) assumption \cite{meadows1986more}. This class of design is a hybrid model. Here both the client and the server need to do some computation on their data set and share the intermittent result to each other through Diffe-Hellman approach, and then perform the PSI, as shown in Figure \ref{fig:diffe}. Here the client's data privacy is protected. This approach is simple with a more computation overhead at the client side, when compared to the hash (na\"ive) based approach, as it needs to compute $O(m+n)$ exponentiations along with $O(mn)$ comparisons to find intersection. For more security, elliptic curve based implementation can also be used.
		
	\subsection{Homomorphic (Additive) Encryption Approach}
\begin{figure}[h]
			\centering \includegraphics[trim=0cm 12.3cm 2.2cm 0cm, clip=true,scale=0.4]{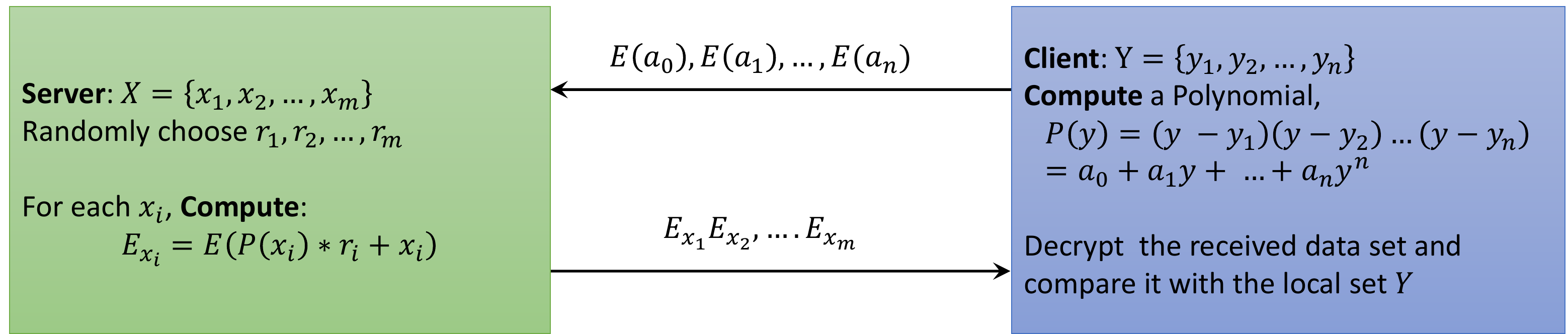} %trim=l b r t
			\caption{PSI Scheme based on Homomorphic Encryption Approach} \label{fig:homomorphic}
			\Description{}
		\end{figure}

		This approach is based on OT technique, which uses additive homomorphic encryption such as Paillier scheme \cite{freedman2004efficient}, shown in Figure \ref{fig:homomorphic}. Here the client constructs a polynomial using its data set and sends the encrypted (using its public key) coefficients to the server. Then server generates a set of random numbers $\lbrace r_1,r_2, \dots, r_m \rbrace$ and for each data $x_i \in X$, it computes $E_{x_i} = E(P(x_i)*r_i + x_i)$ using homomorphic encryption and sends it to the client. Then the client decrypts the received data set and determines the intersection. Here for the server, there is a computation overhead due to polynomial evaluation on the encrypted data and one can explore efficient polynomial evaluation such as Karartsuba method to speed up.
	
	\subsection{Blind RSA Approach}
\begin{figure}[h]
		\centering \includegraphics[trim=0cm 10.6cm 0.5cm 0cm, clip=true,scale=0.35]{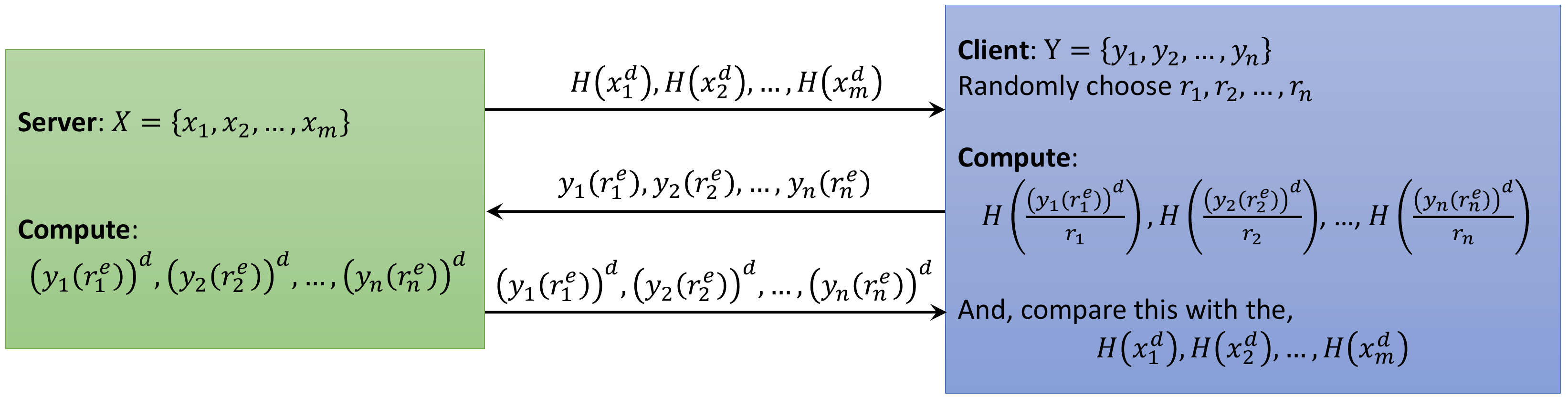} %trim=l b r t
		\caption{PSI Scheme based on Blind RSA Approach} \label{fig:blindrsa}
		\Description{}
	\end{figure}

	In this approach \cite{de2010practical}, security of the algorithm is based on RSA factorization, shown in Figure \ref{fig:blindrsa}. Here the server choses $N=pq,e,d$, where $p,q$  are prime numbers and $e,d$ are public and private keys respectively, generated by the server. Note that $e,d$ are co-prime to $N$ and $ed \equiv 1 mod(N)$. Using RSA technique, both the client and the server perform the computations and finally the client computes PSI. Here the client's data privacy is protected, as the server cannot learn anything about the client's data.
	%\end{itemize}
	
	\section{Complexity Analysis of PSI Schemes}
	In Table \ref{tab:psicomplexity}, we have tabulated the computational and communication complexities for the schemes described earlier (Section \ref{sec:schemes}) by theoretical analysis. We assume that the input data size is $\alpha$ bits, output data size is $\beta$ bits and security parameter is $\tau$ bits. We derived the expressions for complexities of the PSI schemes based on the complexities of primitive operations. These PSI schemes are inherently designed using the primitive operations that are described in the appendix.
	
	\begin{table} [ht]
		\centering
		%\footnotesize
		\begin{tabular}{|p{2.7cm}|p{2.7cm}|p{2.7cm}|p{2cm}|p{2cm}|} 
			\hline
			\multirow{2}{*}{PSI Scheme}             & \multicolumn{2}{c|}{\begin{tabular}[c]{@{}c@{}}Computation Complexity\\(\#primitive operations) \end{tabular}}           & \multicolumn{2}{c|}{\begin{tabular}[c]{@{}c@{}} Communication Complexity \\(in bits) \end{tabular}}   \\ 
			\cline{2-5}
			& \begin{tabular}[c]{@{}c@{}}Server \end{tabular} & \begin{tabular}[c]{@{}c@{}}Client \end{tabular} & Server & \begin{tabular}[c]{@{}c@{}}Client \end{tabular}                                              \\ 
			\hline
			Hash based (Na\"ive) approach (pull model) & $O(m(\alpha + \beta)\tau)$ & $O(n(\alpha + \beta) \tau + mn)$ & $O(1)$ & $O(m\beta)$ \\ \hline
			
			Hash based (Na\"ive) approach (push model) & $O((m(\alpha + \beta)\tau)) + mn$ & $O(n(\alpha + \beta)\tau)$ & $O(|X \cap Y|\beta)$ & $O(n\beta)$ \\ \hline
			
			Diffie-Hellman approach & $O(\tau(m\alpha^2+n\tau^2 ))$ & $O(\tau(n\alpha^2+m\tau^2 )+mn)$ & $O((m+n)\tau)$ & $O(n\tau)$ \\ \hline
			
			Blind RSA approach & $O(\tau(m\alpha^2+4n\tau^2 )) $ & $O(n(\alpha^2\tau + 2\alpha\tau + 2\tau(\log 2\tau )^2 + m))$ & $O(2\tau(m+n))$ & $O(2n\tau)$ \\ \hline

			Homomorphic (Additive) Encryption approach & $O(m(n(4\tau\alpha+16\tau^2 )+4\alpha\tau+16\tau^2+2\tau^2(\tau+(2\alpha+8))))$ & $O(n^2(\alpha^2+\alpha)+n(2\tau^2 (\tau+2\alpha+8)))+m(4\tau(16\tau^2+4\tau+(\log 4\tau)^2)))+mn)$ & $O(4m\tau)$ & $O(4n\tau)$
			\\ \hline
		\end{tabular}
		\caption{\small{Computation and Communication Complexity of PSI Schemes. For \emph{Blind RSA Approach} \& \emph{Homomorphic Encryption Approach}, it is assumed that inputs are hashed and keys are generated. Efforts for hashing and key generations are not included.}} \label{tab:psicomplexity}
	\end{table}

	\section{Computation Complexity of PSI Schemes For Contact Tracing in Indian Scenario}
	For the performance analysis of PSI schemes for contact tracing applications, we consider the India scenario. We know that India's population is about 130 crores (that is, 1.3 billion), which is approximately equal to $2^{31}$. Let us assume that, 50\% ($2^{30}$) of the people have smart phones. Let us also assume that among every 500 persons, one person (that is, 2000 cases per 1 million populations, as observed in hot zones like USA, Italy, and France\footnote{https://www.worldometers.info/coronavirus/}), is infected with COVID-19. Then the total number of affected people with COVID-19 is approximately equal to $2^{22}$. We assume that the contact tracing app collects the GPS trails of affected persons. For the analysis purpose, the app records the GPS trail of a person once in a second. For a month, number of GPS trails related to each person collected is 2592000 trails. Since, a lot of GPS trails related to the person may not change successively for a long duration in a day due to controlled mobility (because of lockdown), eight hour sleep, etc., for the analysis purpose, we sample the data and use only 25\% of the collected data, which is approximately equal to $2^{20}$. In totality, at the server side, there are $2^{42}$ ($=2^{22} 2^{20}$) GPS trails. We also assume that this data is stored town wise in the server database. In India, we have about 7935 towns ($\approx2^{12}$). Therefore, town wise, there can be ($\approx2^{30}$) GPS trails of the COVID-19 affected persons on an average. Hence a person who is residing in a particular town can find out whether he/she was in close proximity to any COVID-19 affected person around a particular time and for how much duration, while ensuring that their privacy is preserved by sharing his/her encrypted GPS trails to the server (push model) or downloading the list of encrypted trails of the affected persons from the server (pull model) of that town. For the performance analysis, we assume that the person using the contact tracing app can send a maximum of $2^{10}$ encrypted GPS trails for the query purpose. We also assume that the server and the user of the mobile phone, where client side of the contact tracing app is running, have CPUs with clock speed of 100 GHz and 1 GHz respectively. 
	
	Table \ref{tab:perf1} and Table \ref{tab:perf2} describe the performance (empirical) analysis of various PSI schemes by evaluating the formulas (computation and communication complexities) described in Table \ref{tab:psicomplexity} for India scenario ($m=2^{30},n=2^{10}$) and also a sparser scenario with $m=2^{20}$ and $n=2^{10}$ respectively.
	
	\begin{itemize}
		\item \textbf{\emph{Server Side Analysis}}. We observe that from the server perspective with respect to the following:
			\begin{itemize}
				\item \textbf{\emph{computation complexity}}. Hash based (na\"ive) pull model is the best scheme followed by hash based (na\"ive) push model, Diffie-Hellman, Blind RSA and Homomorphic (Additive) Encryption approaches.
				
				\item \textbf{\emph{communication complexity}}. Hash based (na\"ive) pull model is the best scheme followed by hash based (na\"ive) push model, Diffie-Hellman, Blind RSA, and Homomorphic (Additive) Encryption approaches.			
			\end{itemize}
		\item \textbf{\emph{Client Side Analysis}}. We observe that from the client perspective with respect to the following:
			\begin{itemize}
				\item \textbf{\emph{computation complexity}}. Hash based (na\"ive) push model is the best scheme followed by hash based (na\"ive) pull model, Blind RSA, Diffie-Hellman and Homomorphic (Additive) Encryption approaches.
				
				\item \textbf{\emph{communication complexity}}. Hash based (na\"ive) push model is the best scheme followed by Diffie-Hellman, Blind RSA, Homomorphic (Additive) Encryption and hash based (na\"ive) pull model approaches.			
			\end{itemize}
	\end{itemize}

	\begin{table}
		\centering
		\begin{tabular}{|p{2.3cm}|p{1.7cm}|p{1.7cm}|p{1.5cm}|p{1.5cm}|p{2.4cm}|} 
			\hline
			\multirow{2}{*}{PSI Scheme}             & \multicolumn{2}{c|}{\begin{tabular}[c]{@{}c@{}}Computation Complexity\\(in seconds) \end{tabular}}           & \multicolumn{2}{c|}{\begin{tabular}[c]{@{}c@{}} Communication Complexity \\(in bits) \end{tabular}} & \multirow{2}{*}{Remarks}  \\ 
			\cline{2-5}
			& \begin{tabular}[c]{@{}c@{}}Server \end{tabular} & \begin{tabular}[c]{@{}c@{}}Client \end{tabular} & Server & \begin{tabular}[c]{@{}c@{}}Client \end{tabular}                                              &                           \\ 
			\hline
			Hash based (Na\"ive) pull model & 2111 & 1099 & $O(1)$  & $2^{38}$& $\alpha=2^9, \beta =2^8, \tau=2^8$\\\hline
			
			Hash based (Na\"ive) push model & 2122 & 0.2 & $2^{18}$  & $2^{18}$& $\alpha=2^9, \beta =2^8, \tau=2^8$\\\hline
			
			Diffie-Hellman & $2.8830 \times 10^6$ & $7.3786 \times 10^{10}$ & $\approx 2^{42}$  & $2^{22}$& We assume inputs are hashed. Input size $\alpha = 2^8 $, Security parameter $\tau = 2^{12}$ \\\hline
			
			Blind RSA & $2.8851 \times 10^6$ & $1377$ & $\approx 2^{43}$  & $2^{23}$& We assume inputs are hashed and keys are generated. Hashing efforts not included. Prime numbers $p$ and $q$ are of $2^{12}$ bits. Input size $\alpha=2^8$, Output size $\beta=2^{24}$ and Security parameter $\tau=2^{12}$\\\hline
			
			Homomorphic (Additive) Encryption & $4.6636 \times 10^9$ & $4.7226 \times 10^{12}$ & $2^{44}$  & $2^{24}$ & We assume inputs are hashed and keys are generated. Prime numbers $p$ and $q$ are of $2^{12}$ bits. Input size $\alpha=2^8$ and Security parameter $\tau=2^{12}$\\\hline
			
		\end{tabular}
		\caption{Performance Analysis of PSI Schemes with $m=2^{30}$ and $n=2^{10}$.} \label{tab:perf1}
	\end{table}

	\begin{table}
		\centering
		\begin{tabular}{|p{2.3cm}|p{1.7cm}|p{1.7cm}|p{1.5cm}|p{1.5cm}|p{2.4cm}|} 
			\hline
			\multirow{2}{*}{PSI Scheme}             & \multicolumn{2}{c|}{\begin{tabular}[c]{@{}c@{}}Computation Complexity\\(in seconds) \end{tabular}}           & \multicolumn{2}{c|}{\begin{tabular}[c]{@{}c@{}} Communication Complexity \\(in bits) \end{tabular}} & \multirow{2}{*}{Remarks}  \\ 
			\cline{2-5}
			& \begin{tabular}[c]{@{}c@{}}Server \end{tabular} & \begin{tabular}[c]{@{}c@{}}Client \end{tabular} & Server & \begin{tabular}[c]{@{}c@{}}Client \end{tabular}                                              &                           \\ 
			\hline
			Hash based (Na\"ive) pull model & 2 & 1.3 & $O(1)$  & $2^{28}$& $\alpha=2^9, \beta =2^8, \tau=2^8$\\\hline
			
			Hash based (Na\"ive) push model & 2 & 0.2 & $2^{18}$  & $2^{18}$& $\alpha=2^9, \beta =2^8, \tau=2^8$\\\hline
			
			Diffie-Hellman & 3518 & $7.2057 \times 10^{7}$ & $\approx 2^{28}$  & $2^{22}$& We assume inputs are hashed. Input size $\alpha = 2^8 $, Security parameter $\tau = 2^{12}$ \\\hline
			
			Blind RSA & 5629 & 279 & $\approx 2^{33}$  & $2^{23}$& We assume inputs are hashed and keys are generated. Hashing efforts not included. Prime numbers $p$ and $q$ are of $2^{12}$ bits. Input size $\alpha=2^8$, Output size $\beta=2^{24}$ and Security parameter $\tau=2^{12}$\\\hline
			
			Homomorphic (Additive) Encryption & $4.5543 \times 10^6$ & $4.6121 \times 10^{9}$ & $2^{34}$  & $2^{24}$ & We assume inputs are hashed and keys are generated. Prime numbers $p$ and $q$ are of $2^{12}$ bits. Input size $\alpha=2^8$ and Security parameter $\tau=2^{12}$\\\hline
			
		\end{tabular}
		\caption{Performance Analysis of PSI Schemes with $m=2^{20}$ and $n=2^{10}$.} \label{tab:perf2}
	\end{table}

	\section{PSI Optimizations Suitable for Mobile Application Deployments}
	There are several optimizations with respect to storage, communication and computation discussed in \cite{kiss2017private,pinkas2018scalable,falk2018private} for PSI schemes. In order to compute intersection of two sets with equal cardinalities (say, $N$); $O(N^2)$ pairwise comparisons need to be performed. By mapping elements of set into hash tables of different bin sizes, intersection of two sets can be computed. Here if elements of the two sets are mapped to the same bin, it implies that they are part of the intersection. In this way the number of pairwise comparisons can be reduced from $O(N^2)$ to $O(N)$. PSI schemes based on OT and Circuit based protocols use these hash table techniques to reduce the complexity of computation. However, there can be false positives in this approach. To reduce the false positives to the desired level, parameters such as the number of bins, the number of hash functions need to be configured appropriately. We discuss the same in the following approaches.
	
	\begin{itemize}
		\item \textbf{\emph{Privacy Enabled Bloom Filter}}. Primarily the two data sets can be mapped into a Bloom filter, which is a compact representation of the set and one can compute the set intersection on respective Bloom filters with approximations. Note that, there can be false positives, but no false negative, when one queries into a Bloom filter. One can reduce the false positives by configuring the Bloom filter parameters such as size, etc. There are privacy enabled Bloom filters that can protect privacy of both the parties (server and client). In \cite{kiss2017private}, privacy enabled Bloom filter based on the above cryptographic approaches is analysed with respect to computing time, storage and communication overheads. This protocol runs in 3 phases: (i) Base phase: Data independent computations are performed. (ii) Setup phase: Data transfer from server to client takes place with O(M) complexity. (iii) Online phase: Computation and data transfer takes place between the server and the client with O(N) complexity. They have implemented the same on both mobile and laptop. On mobile phone, for a maximum cardinality of set $2^{20}$ with 128-bit security, performance of Diffie-Hellman (in online phase) based Bloom filter is approximately 2.5 to 10 times faster when compared to RSA and ECC based Bloom filters.
		
		\item \textbf{\emph{Cuckoo Hashing}}. Cuckoo hashing is a data structure, which uses a set of $\gamma$ hash functions $\lbrace h_1,h_2, \dots, h_\gamma \rbrace$ wherein all the elements of a set can be mapped into different bins without any collision in a compact way. In case of a collision, the current element is relocated into another location. With this, any query for the set membership can be achieved with a constant computation complexity $O(1)$. There are several schemes \cite{pinkas2018scalable,jack2019categorization} that use Cuckoo hashing to design PSI in an efficient manner, both in terms of computation and communication. For example, in \cite{pinkas2015phasing}, it is shown that OT-phasing based PSI protocol requires a factor of 2-3 less communication than that of DH-ECC based PSI protocol.
		
		\item \textbf{\emph{Count Min Sketch/Counting Bloom filter}}. In order to reduce the False Positive Rates (FPR), count min (CM) sketch can be used for PSI in a compact way. According to \cite{cormode2005improved}, \emph{"CM sketch is a probabilistic data structure that serves as a frequency table of events in a stream of data. It uses hash functions to map events to frequencies, but unlike a hash table, uses only sub-linear space, at the expense of over counting some events due to collisions"}. This needs to be explored from privacy preserving computation, communication, storage and update perspective. The advantage with this data structure is besides computing PSI, one can also record/count the frequency of the data element that is repeated in the set (multi-set) in a compact way. For example, the number of COVID-19 positives in a particular location spotted where a normal healthy person visited at the same time can be computed along with the matching in a compact way.
	\end{itemize}

	\section{PSI Schemes Used in Contact Tracing Applications}
	In general, a contact tracing application can be designed to operate in a centralised, decentralised or hybrid manner. In the centralised design approach, data belonging to the application user is managed by the centralised server. As a result of this, the server is in a position to carry out surveillance. For instance, apps such as Hamagen (The Shield) from Israel, China app and Aarogya Setu of India are based on the centralized design. Further, these apps do not use any PSI schemes as of writing this paper \cite{shukla2020privacy}.
	
	When such apps are designed with a decentralised approach, the application users have a greater control on their data. Only those users who are confirmed to be infected with COVID-19 may voluntarily share their data (such as GPS trails) to certain listed entities like government, health care agencies, etc., in a privacy preserving manner. In \cite{troncoso2020decentralized}, DP3T (Decentralised Privacy Preserving Proximity Tracing) architecture is proposed that is based on a decentralised design. To enable privacy and security of the user, they have proposed secret sharing schemes and pseudonyms. No information on the use of PSI scheme is discussed in the paper. 
	
	Applications such as TraceTogether from Singapore follow the hybrid design paradigm. Here anonymity of the application user is assured by a centralised entity (government). When an application user is confirmed to be infected with COVID-19, he/she can voluntarily share the collected data (which is anonymised) to the server. This helps the server to alert the other users (and may even be non-users) who were in the proximity of this user. Since, the centralised server facilitates the data anonymisation for the application users; it can learn information about those users who had uploaded the data. In TraceTogether application, no PSI scheme is used.
	
	The initial version of the MIT's Safepath \cite{raskar2020apps} is based on the centralised design (pull model), in which the server manages the data (GPS trails) related to the COVID-19 affected persons in privacy preserving manner. It facilitates the user of the application to know, whether he/she was in close proximity with the affected person in a privacy preserving way. To enable this, they have implemented PSI scheme based on hash (na\"ive) approach. For the next version of the app, they are coming up with other sophisticated PSI schemes such as Diffie-Hellman based PSI.
	
	\section{Concluding Remarks}
	One of the ways to contain the COVID-19 infection spreading is through contact tracing. However, in some of the countries, contact tracing is done manually, which is time-consuming and cumbersome. Hence, digital contact tracing, which is simple and fast, is preferable over manual contact tracing. The success of the digital contact tracing depends on the number of people using it effectively. One of the hindrances to adopting the digital contact tracing by the citizens is privacy concerns. Digital contact tracing without privacy of the users can lead to their surveillance by the agencies such as the government. Most of the existing contact tracing applications have limited data privacy for their users. Hence, privacy enabled contact tracing applications is the need of the hour. In this direction, we have discussed some of the PSI schemes that can be leveraged to enable data privacy for the user of the contact tracing applications. Further, from both the user's and the server's perspective, we analysed the security, privacy and performance of these schemes for the contact tracing application. We also analysed empirically the performance of these PSI schemes by taking into consideration the scenario in India. As part of the future work, we would like to explore the scope of optimization of the PSI schemes for designing efficient digital contact tracing applications.

\bibliographystyle{ACM-Reference-Format}
\bibliography{psidesignapproaches}

\appendix

	\section{Appendix}
	The computational complexities of the various operations discussed in this section are referred from \cite{knuth1981art,paillier1999public,preneel1993analysis}.
	
	\begin{enumerate}
		\item Calculation for number of digits of GPS data and time stamp
			\begin{itemize}
				\item Latitude = 10 digits = 10 bytes
				\item Longitude = 10 digits = 10 bytes
				\item Time (long) = 8 bytes = 16 digits = 16 bytes
				\item Concatenation = 36 bytes $\approx2^9$ bits  				
			\end{itemize}
		\item Computation Complexity of hash function 
			\begin{itemize}
				\item Depends on
				\begin{itemize}
					\item input size in bits ($\alpha$)
					\item output size in bits ($\beta$)
					\item security parameter in bits ($\tau$)			
				\end{itemize}
				\item Computational complexity is $O((\alpha+\beta)\tau)$ 
				\item For contact tracing, $\alpha=2^9$, $\beta=2^8$ and $\tau=2^8$, the time complexity for one GPS data is $O(3\times2^{16})\approx 1.96608\times10^5$ instructions
				\item On 100 GHz machine, it takes approximately around 1.9661 micro seconds to compute one hash value of a GPS data
				\item On 100 GHz machine,  to hash 100 billion GPS data, it takes approximately around  $1.96608\times10^5$ seconds							
			\end{itemize}
		\item Computation Complexity of addition/subtraction of two numbers
		\begin{itemize}
			\item Depends on sizes of the two numbers. We assume that both the numbers are of $\alpha$ bits length
			\item Computational complexity is $O(\alpha)$						
		\end{itemize}
		\item Computation Complexity of multiplication of two numbers
		\begin{itemize}
			\item Depends on sizes of the two numbers. We assume both the numbers are of $\alpha$ bits length
			\item Computational complexity is $O(\alpha^2)$			
		\end{itemize}
		\item Computation Complexity of division of two numbers
		\begin{itemize}
			\item Depends on sizes of the two numbers. We assume both the numbers are of $\alpha$ bits length
			\item Computational complexity is $O(\alpha(\log \alpha)^2)$	
		\end{itemize}
		\item Computation Complexity of modular exponentiation
		\begin{itemize}
			\item Depends on sizes of the base ($\alpha$) and exponent ($k$)
			\item Computational complexity is $O(\alpha^2 k)$	
		\end{itemize}
		\item Computation Complexity of polynomial evaluation
		\begin{itemize}
			\item Depends on  the degree of the polynomial, size of the co-efficient , number of variates (variables) and size of the variable						
			\item We assume degree of polynomial is $\gamma$ and size of the coefficients and variable is $\alpha$
			\item Computational complexity (using Horner's method) is $O(\gamma\alpha(\alpha+1))$			
		\end{itemize}
		\item Computation Complexity to find the coefficients of a polynomial from its roots
		\begin{itemize}
			\item Depends on the degree of the polynomial (number of roots), size of the roots
			\item Using fundamental theorem of algebra (Vi\"ete's formula), coefficients of a polynomial from its roots can be determined
			\item We assume degree of polynomial is $\gamma$ and size of the roots and variable is $\alpha$
			\item Computational complexity is $O(\gamma^2 (\alpha^2+\alpha))$					
		\end{itemize}
		\item Computation Complexity of Paillier additive homomorphic encryption	
		\begin{itemize}
			\item As part of the setup, $u=pq$, where $p,q$ are two large prime numbers with $gcd(pq,(p-1)(p-1))=1$
			\item $\aleph=lcm(p-1,q-1)$, $g \in Z_{u^2}^*$ and $\mu = (L(g^\aleph(mod\text{ }u^2)))^{-1} mod\text{ }u$, where $L(\theta)=(\theta - 1)/u$
			\item Public key is $(u,g)$ and Private key is $(\aleph,\mu)$
			\item Encryption of message $s$ is $c = E(s) = g^s r^u mod\text{ }(u^2)$, $r$ is random and $gcd(r,u)=1$
			\item Decryption of message is $s = D(c) = L(c^\aleph mod\text{ }u^2)\mu\text{ }mod\text{ }(u)$
			\item Homomorphic addition of two messages $s_1$ and $s_2$ is $s_1 + s_2 = D(E(s_1,r_1 )*E(s_2,r_2 ))$
			\item We assume that $p,q$ are of size of $\alpha$ bits and thus $g,r$ are of $2\alpha$ and $\alpha$ bits respectively, message is of $\beta$ bits 
			\item For complexity analysis, we assume that public and private keys are generated
			\item Computational complexity of encryption, decryption and homomorphic addition are $O(2\alpha^2(\alpha+2\beta+8))$, $O(4\alpha(16\alpha^2+4\alpha+(\log 4\alpha)^2))$ and $O(16\alpha^2)$ respectively
								
		\end{itemize}
	\end{enumerate}

\end{document}